# Do recommender systems function in the health domain: a system review

Jia Su[a], Yi Guan[a], Yuge Li[a], Weile Chen[a], He Lv[b], Yageng Yan[b]

[a] School of Computer Science and Technology, Harbin Institute of Technology, No. 92 West Dazhi Street, Harbin, China

[b] Nutritional Department, The 1st Affiliated Hospital of Harbin Medical University, No. 23 Youzheng Street, Harbin, China

**Abstract**

Recommender systems have fulfilled an important role in everyday life. Recommendations such as news by Google, videos by Netflix, goods by e-commerce providers, etc. have heavily changed everyone's lifestyle. Health domains contain similar decision-making problems such as what to eat, how to exercise, and what is the proper medicine for a patient. Recently, studies focused on recommender systems to solve health problems have attracted attention. In this paper, we review aspects of health recommender systems including interests, methods, evaluation, future challenges and trend issues. We find that 1) health recommender systems have their own health concern limitations that cause them to focus on less-risky recommendations such as diet recommendation; 2) traditional recommender methods such as content-based and collaborative filtering methods can hardly handle health constraints, but knowledge-based methods function more than ever; 3) evaluating a health recommendation is more complicated than evaluating a commercial one because multiple dimensions in addition to accuracy should be considered. Recommender systems can function well in the health domain after the solution of several key problems. Our work is a systematic review of health recommender system studies, we show current conditions and future directions. It is believed that this review will help domain researchers and promote health recommender systems to the next step.

**Abbreviations**

HRS: health recommender system
RS: recommender system
IF: information filtering
CB: content-based filtering
CF: collaborative filtering
CKD: chronic kidney disease
KB: knowledge-based
SVD: singular value decomposition
WHO: World Health Organization
NCDs: noncommunicable diseases
BMR: basal metabolic rate
FSA: Food Standards Agency

BMI: body mass index
GPS: Global Position System
GSM: Global System for Mobile Communications
SWRL: semantic web rule language
JESS: Java expert system shell
CBR: case-based reasoning
IoT: Internet of Things
CVD: cardiovascular disease
BN: Bayesian network
RF: random forest
PHR: personal health record
IR: information retrieval
TF-IDF: term frequency/inverse document frequency
SGD: stochastic gradient descent
ICR: insulin to carbohydrate ratio
ISF: insulin sensitivity factor
OWL web ontology language
API: application program interface
RDF: resource description framework
MAE: mean absolute error
RMSE: root of mean square error
ROC: receiver operating characteristic
NDCG: normalized cumulative discounted gain
DCG: discounted cumulative gain

## 1. Introduction

A health recommender system (HRS) is a specialized application of a recommender system (RS) that refers to using an RS to solve health decision-making problems in scenarios such as diet [1-6], health education [7-10], activity [11-14], medical progress [15-18], etc. RSs are software tools and techniques that provide suggestions for items to be of use to a user [19]. In detail, the "tools" are always represented by a user-centered system with an interaction interface, background suggestion model, network, and database. "Techniques" typically refers to an information filtering (IF) method, the most common content-based filtering (CB), collaborative filtering (CF), and hybrid filtering. However, the "to be of use" is always measured such as by a rating score to quantify the usefulness of an "item" (e.g., movie, or book)[20,21]. RSs have achieved great success in commercialized scenarios; these account for 2/3 of all Netflix movies watched and 35% of Amazon.com sales [22-29]. However, health RSs are still in their infancy due to the inherently complex nature of health. The earliest article searched from Google Scholar is Pattaraintakorn et al. [30], which was published in 2007 with the query "health recommender system". Concerning healthcare, a minor decision may have butterfly effects; i.e., the potential risks under a decision are hardly predictable. All the "tools", "techniques", "items" and "to be of use" elements should be carefully reconsidered. Thus, a conservative and heuristic principle is popular. The finding in our survey process also verifies this principle that lifestyle, especially diet HRSs occupies more studies than do medicine, diagnosis and other medical recommendations, as the risk of diet seems to be the least. This finding is a trade-off between health domain challenges and reality.

The challenge of HRSs is multi-faceted [31], and can be summarized with the following items: complexity and specification, end-user diversity, usage scenario variation, data availability including user profile, user-item rating, health recommendation items, feedback, and the most difficult evaluation problems. Let's first discuss the items. The items range from diet activities to medical entities; furthermore, each item still has detailed categories. Diet can be subdivided into food, recipe, grocery, and menu recommendations. Additionally, when changing food into a recipe scenario, the rationality of a food match should be concerned; for a menu scenario, nutritional constraints also should be kept in mind. For end-users, the common users are patients, doctors, and nurses who can obtain imagined benefits such as lifestyle, diagnosis or disposal recommendations. In fact, healthy individuals, researchers and policy-maker, etc., can obtain benefits [31]. The usage scenarios are varied; for instance, similarly to the diet RS, a system can be set to different goals such as weight loss, muscle gain, and blood pressure reduction. For different scenarios, the HRS has to change its recommendation strategy to adapt to the user's actual demand. Unlike RSs in other domains in which benchmark datasets exist, such as MovieLens [32], Netflix [28], and Jester [33], for health, most research datasets are built from scratch. Freyne and Berkovsky [1] made food recommendations with a self-collected dataset of 183 users rating 20 foods or 30 recipes and developed a scale of 8701 preferences. Chi et al. [34] proposed a diet consultation system for chronic kidney disease (CKD) patients that can be regarded as a diet RS; they collected common CKD knowledge, controlled vocabulary, and 84 patient records from a hospital. The evaluation of an HRS is always regarded as a multidimensional procedure. For RSs in traditional domains, user satisfaction is the first principle, but not for HRSs. There is consensus that safety should always be the first consideration [3,35,36] for health. Additionally, privacy, precision, consequence, serendipity, coverage, individuation, satisfaction, and trust [37-44] all are criteria to assess an HRS. All of these challenges make HRS a difficult direction to pursue in the RS scientific research and commerce fields.

In this work, we want to clarify the current status and future direction of HRSs. We review the studies and related works and hope this review may help the development of HRS studies and applications. This paper is structured as follows. In section 2, we list main interests in HRSs. Section 3 states the algorithms, methods and datasets in HRSs. Section 4 discusses the evaluation problem. Section 5 discusses challenges and trends, and section 6 shows the conclusion.

## 2. Interests in health recommender system

The topics in HRSs, learned from previous review works such as Valdez et al. [31] and Wiesner and Pfeifer [45], etc., are mainly focus on lifestyle and medically relevant information recommendations. Using the Google Scholar website, we investigated research papers published in the last decade for related terms. The related terms we used are the following: 1) diet recommender system (food, meal, and recipe); 2) physical activity recommender system (exercise); 3) lifestyle recommender system; 4) medical recommender system (healthcare, medicine, drug, and diagnosis); and 5) health education recommender system. Furthermore, not only these retrieved papers but also ones with semantic relations (i.e., those that refer to terms cited in the retrieved paper) were also considered. For highlighting key studies and trends, only papers with more than 50 citations or published on line in the most recent 5 years were selected in this review. The number of carefully selected articles is 38, and the distribution of topics is shown in Fig. 1. We list the main features of each article with algorithm, items, end-user, data, evaluation and keywords in Table 1. The food recommendations are divided into two parts: preference-based and constraint-based.

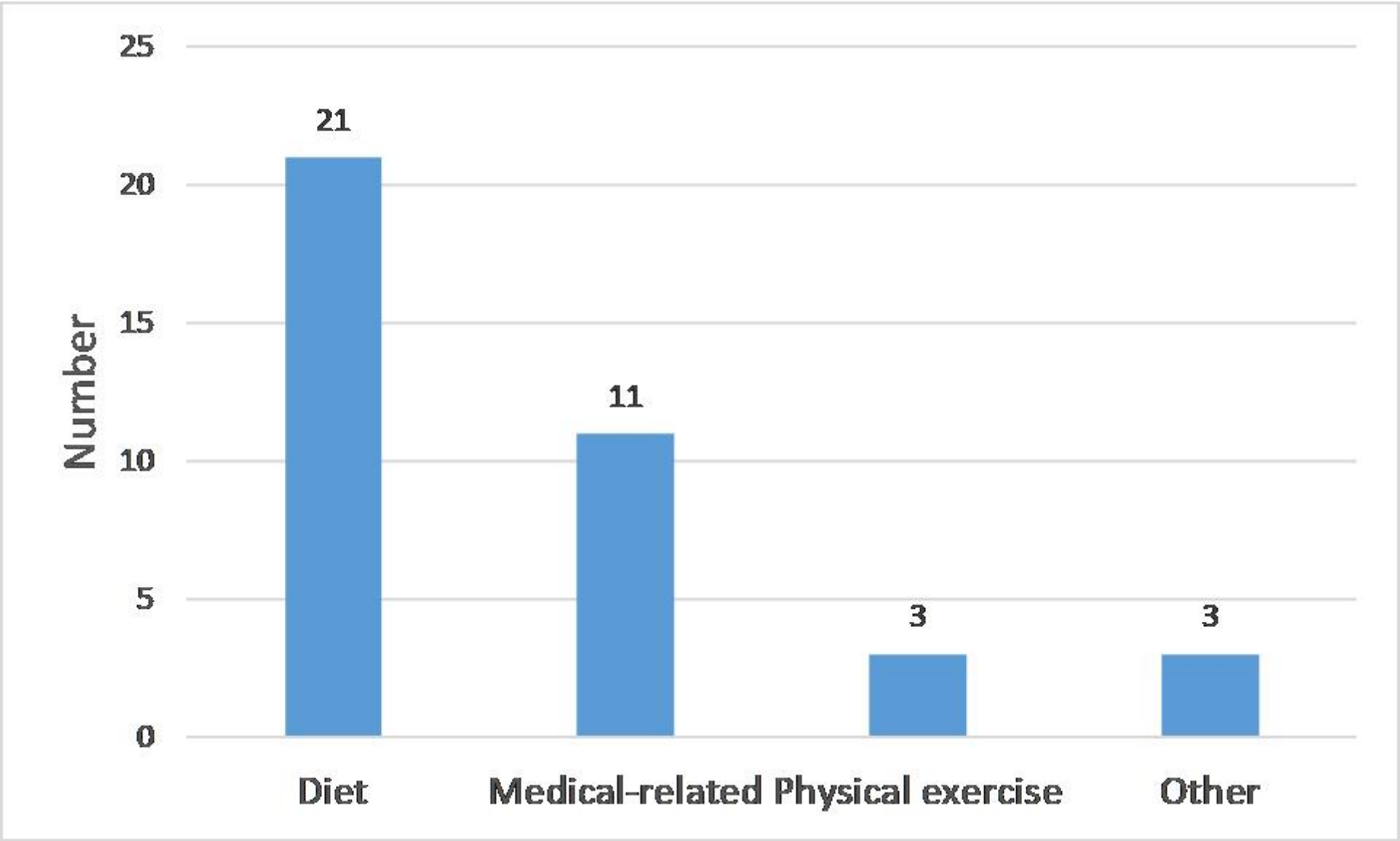

Fig. 1 The interest topic distributions of articles in this discussion

Table 1. HRS studies in the last decade

| HRS | Paper (year, citations) | Algorithm | Items | End-user | Data | Evaluation | Keywords |
|---|---|---|---|---|---|---|---|
| Diet (preference-based) | Ornab et al [46] (2019, 0) | CF | Food | Single user | - | Accuracy (Top-ten consistency) | cosine similarities, alternating least squares |
| | Achananuparp and Weber [47] (2016, 10) | CB | Food | - | MyFitnessPal | Accuracy | Food substitutes, distributional similarity |
| | Ge et al [48] (2015, 32) | Hybrid | Recipe | Single user | Total Wellbeing Diet | Accuracy, usability | User tags |
| | Teng et al [6] (2012, 173) | CF | Recipe | Single user | Allrecipes.com | Accuracy | Ingredient networks |
| | Forbes and Zhu [49] (2011, 112) | CF | Recipe | Single user | Allrecipes.com | Accuracy | Matrix factorization |
| | Pinxteren et al [50] (2011, 75) | CB | Recipe | Single user | Dutch recipe website | Effectiveness, satisfaction | Recipe similarity |
| | Ueda et al [51] (2011, 58) | CB | Recipe | Single user | Cookpad | - | Extraction ingredient |
| | Berkovsky and Freyne [52](2010, 202) | CF | Recipe | Group user | CSIRO Total Wellbeing Diet book | Accuracy, Coverage | Family, user weighting |
| | Freyne and Berkovsky [1] (2010, 131) | CB, CF, Hybrid | Recipe/Food | Single user | CSIRO Total Wellbeing Diet book | Accuracy, Coverage | Personalization |
| Diet (constraint-based) | Agapito rt al [53] (2018, 8) | KB | Food | Single CKD, diabetes, hypertension | ICD9-CM, Calabrian foods | - | Questionnaires |
| | Trattner and Elsweiler [54] (2017, 30) | KB, CF | Recipes, Meal plan | Single user | Allrecipes.com, WHO, FSA | Healthiness, Accuracy | Evaluation |
| | Yang et al [55] (2017, 24) | CF, KB | Recipe | Single user | Yummly API, Food-101 | Acceptance | Image similarity, survey |
| | Elsweiler et al [56] (2017, 22) | CF, KB | Recipe | Single user | Allrecipes.com | Accuracy | Choice biases, healthier |
| | Bianchini et al [57] (2017, 15) | CB, KB | Menu | Single user | BigOven.com | Response time, quality | Prescription, menu generation |
| | Rehman et al [58] (2017, 3) | Demographic-based, CB | Food item list | Single patient | Laborities, Official website | Accuracy | Nutrition requirement |
| | Chen et al [59] (2017, 1) | KB | Food | Single CKD | Taiwan food nutrition database | Accuracy | Jena |
| | Jung and Chung [60] (2016, 64) | KB, CF | Menu | Single obesity | SeeMe5 | Accuracy, Satisfaction | Knowledge |
| | Chi et al [34] (2015, 30) | KB | Food group, Nutrient | Single CKD | Common CKD knowledge, government open data sets | Accuracy | Semantic rules |
| | Elsweiler and Harvey [61] (2015, 27) | Demographic-based, CF | Meal plan | Single user | NHANES | Accuracy | Nutrition |
| | Lee et al [62] (2010, 203) | KB, fuzzy set | Menu | Single diabetes | Taiwanese food ingredients database | Satisfaction | Imprecise and vague knowledge |
| | Phanich et al [63] (2010, 50) | CB, KB | Dish | Single diabetes | Nutritive values for Thaifood | Acceptance | Nutrition, Nutritionist |
| Physical activity | Costa et al [64] (2018, 10) | CF, Glocko2 rating algorithm | Exercise | Single elder | British National Health Security | - | Exercise recognition |
| | Agu and Claypool [29] (2016, 4) | CF | Exercise game | Single user | Funf | - | Questionnaire, E-score |
| | Bielik et al [14] (2012, 55) | KB | Training plan | Single/Group user(s) | - | Understanding | Motivation |
| Medically related | Jabeen et al [65] (2019, 0) | CF, demographic-based | CVD types and advice | Single user | Hospital | Accuracy (Precision, Recall, MAE) | IoT, CVD classification |
| | Chiu and Chen [66] (2019, 0) | Fuzzy learning | Clinic | Single patient | - | Accuracy | Successful recommendation rate |
| | Jamshidi et al [67] (2018, -) | KB, CF | Health prediction | Single user | Dental, medical, hospital records | Accuracy (Precision, recall) | EHR, ensemble |
| | Torrent-Fontbona and López [17] (2018, 7) | KB (Case-based reasoning) | Bolus insulin | Single Type 1 diabetes | - | Effectiveness, Robustness | ICR, similarity |
| | Zhang et al [68] (2017, 104) | CF, matrix | Doctor | Single user | Yelp | Accuracy | Personalized, Emotional offset, |
| | | factorization | | | | | matrix factorization |
| | Gräßer et al [69] (2017, 9) | CF, Demographic-based | Therapy | Single skin disease patient | Health records from Hospital | Accuracy | Similarity, sparsity |
| | Wang et al [70] (2016, 27) | CF | Health information | Single user | online | Satisfaction, usefulness, value, trust | Cloud computing |
| | Kondylakis et al [71] (2015, 20) | CB | Medical information | Single patient | Online web resources (web pages, pdfs, docs) | Usability, usefulness | Semantic similarity |
| | Zhang et al [72] (2014, 52) | CF, tensor decomposition | Drug | Single user | Online drug store Walgreens | Accuracy (Precision, Recall, F1) | Tensor modeling |
| | Chen et al [16] (2012, 124) | KB | Anti-diabetes drug | Doctor | AACEMG | Accuracy | SWRL, JESS |
| | Duan et al [73] (2011, 220) | CF | Nursing care plan | Single patient | Hospital | Average ranking | Support, confidence |
| Other | Bocanegra et al [74] (2017, 17) | CB | Health educational website | Single user | MedlinePlus, YouTube, SNOMED-CT | Accuracy | Semantic Retrieve |
| | Hors-Fraile et al [75] (2016, 5) | Hybrid (CB, utility-based, CF) | Reminder Message | Single smoker | Google Fit activity data | Accuracy | Message type, Sending time |
| | S. Bauer et al [76] (2012, 102) | KB | Sleep hygiene message | Single user | Sleep literature | Awareness | Peripheral display |

* Note: The year in title "Paper (year, citations)" refers to the publishing year of the paper. The citations refer to the paper citation frequency and are obtained from Google Scholar search (Up to 6/30/2019). KB denotes knowledge-based method.

### 2.1 Diet

2.1.1 *Preference-based*

Preference-based diet recommendations are similar to, for example, commercial recommendations from Amazon for products [24] or from MovieLens for movies [25]. The primary principle in this recommendation is to simultaneously give the user the most preferable diets to replace a monotonous diet and to enrich their diet diversity. The main strategy is to apply the user's ratings, preferences, browsing, cooking, etc. historical data to model the user's interest profile and then to predict the user's preference for new items based on algorithms. For the diet recommendation, ingredients, foods, recipes, menus, and meal plans were frequently divided and assembled for preference model transformation, which we called a division and combination strategy. Freyne and Berkovsky [1] utilized the preferences and ratings of users on recipes and foods to simulate predicting new recipes. They showed that recipe ratings can be transferred into ratings of component foods, and vice versa. Ueda et al. [51] use one's recipe browsing and cooking history to score one's preferences and dislike for ingredients. Then, new recipe scores were generated using the ingredients. Teng et al. [6] constructed two types of networks to capture the interrelations between ingredients: the complement network, which captures the co-occurrence of ingredients in a recipe and the substitute network, which is derived from user-generated suggestions for modifications. The recipe recommendation was based on the ingredient features from the two networks. Ge et al [48] used tags as ingredients or features on recipes. It is inferred that the tags assigned to an item are correlated with the user's rating of a recipe and can be used to better predict the user's interests.

Family-based dietary style has some popularity around the world, and how to quantify and model the influence of family member preference in a group-based food recommendation is presented in [52]. They mainly compared two strategies: the aggregated model strategy, in which family ratings were generated by aggregating individual ratings, and then family predictions were generated using a CF algorithm; and the aggregated prediction strategy, in which individual predictions for a user on an unrated item was first generated by the CF method, and then family predictions were generated by aggregating individuals' predictions. Additionally, the different roles of the individuals in a family was also modeled by a weight strategy to show their influences in a group-based recommendation.

2.1.2 *Constraint-based*

When a recommender system must meet specific requirements, constraint-based recommendations show their importance [77-79]. Diet is not an unconstrained, completely preference-based affair for users with specific diseases [80-84] or merely for nutritional purposes[85-87]. The World Health Organization (WHO) points out that the determinant role of diet for chronic noncommunicable diseases (NCDs) is well established and that diet occupies a prominent position in prevention [83]. The most important thing in a constraint-based diet RS is how to deal with these domain constraints (e.g., expert knowledge, domain limitations, guidelines, and experience, etc.) in the recommending process.

A large interest of diet RSs is the background consideration of diseases such as diabetes, CKD and obesity, etc., which are heavily affected by diet. For diabetes patients, diet nutrition is strictly restricted, but dietary diversity needs must also be met. Phanich et al. [63] searched the similar dishes in nutrition and suggested dishes with nutritionist assistance. They utilized eighteen nutrient values of food, such as energy, water, protein, etc. to categorize and group 290 Thai local dishes using cluster algorithms. Additionally, their dishes dataset was divided into normal food, limited food, and avoidable food based on nutrient recommendations for diabetes by a nutritionist. A recommendation was conducted in each clustered sub-dataset and dish with the least distance to retrieved dish and tagging with normal food or limited food was suggested. When handling the health domain complexity in depth, a fuzzy set may show its ability. Lee et al. [62] applied type-2 fuzzy sets to handle the uncertainties of diabetes patient measurements, changing behavior, exercise patterns, context of the patient, linguistic description, intra-user with varied expert opinions and inter-user with differences among various experts. Additionally, type-2 fuzzy-based personal profile ontology (data from by personal profiles), food ontology (data from by nutrition facts of foods), and personal food ontology (data from acquired diet goal and meal records) were built, and an inference mechanism was implemented to recommend meal allowances. Then, a type-2 fuzzy set based menu-recommendation mechanism recommended some balanced, diverse dinner menus for individual diabetes. For other diseases, Chi et al. [34] constructed a domain ontology to involve controlled vocabulary (activity level, calories level and food groups), CKD knowledge (dietary servings, nutrient limitation, and CKD stage definition) and nutrient compositions with the help of domain experts. Chen et al. [59] used a decision tree to categorize CKD patients into the group category with features of the user's needed nutrient intakes. Combined with nutrient amendment by a nutrition expert system based on the patient's past dietary records and food ontology, a rule-based inference algorithm was conducted for recommendations. Agapitoet al. [53] utilized domain knowledge to categorize diabetes, hypertension and CKD status, and related restrictions were applied in the food recommendations. Jung and Chung [60] focused on diet recommendations for obese adolescents. Domain knowledge including dietary intake data, active mass data, dietary nutrition content data, and exercise content data, etc. were constructed.

Another interest is that of integrating nutritional or health constraints into a normal individual's diet recommendation. Kim et al. [88] thought nutrient requirements could be calculated via empirical equations for individuals and used as constraints in the food recommendation. Elsweiler and Harvey [61] used the Harris-Benedict equation to estimate an individual's basal metabolic rate (BMR) and daily calorie requirements. Moreover, nutritional scientists established generally accepted ranges as guidelines for nutrients. Additionally, the user's profile was built to recommend a preferred diet. In Bianchini et al. [57], ontology-based food and health-related knowledge were linked, and a rule-based menu generation algorithm was utilized. Elsweiler et al. [56] analyzed how to recommend healthier substitute meals using Internet source data. They collected online recipes and established recipe pairs based on a pairwise similarity algorithm. According to the distribution of various health properties across pairs, healthier replacements can be found. Yang et al. [55] considered the problem of combining nutritional expectations with individual tastes in food recommendations. Individual diet type and health goals were collected and fulfilling food recommendations were made with the help of food image analysis. Trattner and Elsweiler [54] utilized WHO guidelines and the United Kingdom Food Standards Agency (FSA) to measure the healthiness of recipes from Allrecipes.com, which is an online recipe sharing and communication platform. Additionally, the most popular recommender algorithms such as Bayesian personalized ranking [89], sparse linear methods [90], and weighted matrix factorization [91], etc. were performed on the rating data. Other explorations not listed in Table 1 include Elsweiler et al. [3],

Harvey and Elsweiler [92], Ueta et al. [93], and Ge et al. [94], which all incorporated healthy problems into the diet RS.

From the above studies, we can see that preference-based and constraint-based diet RSs have essential differences. No matter the diet RS for diseases or normal recommendations with nutritional needs, personal diet constraints based on domain (or expert) knowledge is always the kernel. Abundant food and recipe knowledge databases including nutrition, disease, and health conditions are necessary resources. Individual preferences rank last, but they should not be ignored when designing a person-friendly RS.

### *2.2 Physical Exercise*

Physical exercise recommendations seem to attract less attention than do diet recommendations. For recommending tailored activity items, an individual's physical activity data collection would be essential. Unlike diet feedback and personal diet data collection progress, physical activity variety is always related to time, context, strength, and the individual. Diet recommendation is in vogue for its easy accessibility of diet data and facility of personal data collection, unlike physical exercise. Thus, smart end devices will play an important role [95]. Costa et al. [64] utilized visual technology to recognize elderly people's positions in real time, verifying whether they were exercising correctly. Their technology is a human-shaped robot and receives high acceptance from the elderly. Additionally, the user's visual performance is collected and included as a restriction on rating and ranking suitable exercise recommendations. Agu and Claypool [29] use a smart phone to gather the user's enjoyment of an exergame, which refers to a combination of an exercise and a game. The new exergame will be recommended according to previous game engagement and exercise questionnaires.

Ontology and domain knowledge are also useful tools for exercise recommendations. In Faiz et al. [12], healthy diet and proper exercise with respect to different domains such as gender, weight, age, nutrition values, food preference and exercise were normalized. Then, semantic rules and a reasoning-based approach were used to generate diet and exercise suggestions for diabetes. Pramono et al. [96] also considered constructing an exercise ontology to model the exercise-aware knowledge. The aspects of diabetes, age, BMI (body mass index), blood sugar levels, exercise, intensity, frequency, and duration of physical activity were constructed. Additionally, ontology-based similarity was used to evaluate diabetic patients' similarity, and recommendations were made for similar patients.

The exercise recommendations need to consider a user's willingness [12], and Bielik et al. [14] realized that a physical exercise-encouraging approach should be developed. Mobile phones with various sensors such as accelerometer, GPS (Global Position System), and GSM (Global System for Mobile Communications) were used to collect user data. Key factors such as giving the user proper credit for activity, ensuring fair play, and providing a variety of motivational tools, etc., should be ensured. Self-monitoring, daily targets, goal setting, competition, gamified rewards, and avatar strategies would be designed and joined in the physical activity recommendation.

### *2.3 Medical-related*

In medical RSs, interests range from therapy plan [69,73], medicine [16,17,72], diagnosis [65], health information [70,71], and health prediction [97] to doctor [68] and clinic [66].

Gräßer et al. [69] presented a system consisting of a CF algorithm that takes therapies as items and therapy response as a user's preference. Duan et al. [73] utilized the correlations among nursing diagnoses, outcomes and interventions to create an RS for nursing care plans. A care plan with an items list measured by traditional association rule-like support and confidence was recommended.

Chen et al. [16] constructed an ontology knowledge base with drug nature attributes, type of dispensing, side effects, and patient symptoms. Then, a semantic rule-based language called SWRL (semantic web rule language) and a rule inference engine call JESS (Java expert system shell) were applied for antidiabetic drug recommendations. Torrent-Fontbona and López [17] used a case-based reasoning (CBR) approach to solve the problem of bolus recommendation for type 1 diabetes mellitus. CBR is an approach to solve a new problem by remembering a previous, similar situation and by reusing information and knowledge of that situation [98]. Zhang et al. [72] proposed an online drug recommendation system that can recommend the top-N related medicines for users according to symptom.

Jabeen et al. [65] developed an Internet of Things (IoT)-based CVD (cardiovascular disease) RS to solve the heart disease diagnosis problem based on data collected by using remote bio sensors. Data were extracted using sequential forward selection based on a greed selection strategy, and eight disease categories were targeted for classification.

Wang et al. [70] proposed a system based on a cloud computing environment, integrating mobile communication technology, context-aware technology and wireless sensor networks to build a service for a personalized health information recommendation. Kondylakis et al. [71] provided patients with personalized, condition-related health information recommendations. A patient preference database, medical condition database, term indexing database, and rule database were constructed and used for a recommendation service.

Jamshidi et al. [67] utilized personal health-related data to provide timely and personalized suggestions, aiming to improve consumer health outcomes and prevent complications, then realized cost savings. They used an ensemble method consisting of a probabilistic graphical model as a Bayesian network (BN) that was used to capture the propagation of effects, CF to measure the similarity of situations, and random forest (RF) to generate condition predictions such as a high probability of hypertension, with a recommendation that the user should to visit physician.

Zhang et al. [68] developed a novel healthcare doctor recommendation system. They considered the problem that crowd-sourced reviews were often interfered with by user emotions and proposed a sentiment analysis-based emotion-aware approach to identify emotional offset. Additionally, a topic model used to discover user preference distribution and doctor feature distribution was incorporated into the matrix factorization model to provide more accuracy and personalized recommendations. Chiu and Chen [66] focus on recommending a clinic to the patient according to mobile data on the individual's location, hospital department, and preference. To solve the problem of data deficiency, they modeled the improvement in the successful recommendation rate of clinic recommendations with a learning process.

### *2.4 Others*

The purpose of health education is to promote, maintain or restore the health of individuals by informing and persuading [7]. Educational recommendations play a vital role in augmenting user knowledge, and the tendency toward a better lifestyle, resulting in less disease risk probability and in cost saving. Bocanegra et al. [74] focused on providing online videos on YouTube, with additional educational websites. Semantic level SNOMED-CT terms were extracted from the video title. They crossmatched SNOMED-CT title terms with the terms from the Bio-ontology API to find synonymous MedlinePlus terms; this approach allows them to obtain a web link from the MedlinePlus property. Hors-Fraile et al. [75] promoted an RS for the cessation of smoking. The recommendation mainly refers to motivational messages such as general motivation, diet tips, exercise and active life recommendations, personal physical activity level (instances such as "Hello Peter! You did great yesterday! You were 15 min over your average activity time. Keep up the good work today!"), and tobacco smoking consequences (instances such as "People who smoke develop Reinke's edema a liquid retention in their necks more easily. It's great that you don't smoke anymore! Kind regards, Dr. Laura Carrasco"). The recommendation strategy was a hybrid weighted RS combined with three individual algorithms: demographic filtering, utility-based, and CB. They also considered a message sending a timed RS algorithm to avoid the "robotic" automatic feeling toward the patient based on the patient's actual reading time. Bauer et al. [76] provided a peripheral display on the wallpaper of the user's mobile phone to promote awareness of the impact of activities on sleeping. The recommendations concern the activities such as caffeine consumption, napping, exercising, heavy meal eating, alcohol consumption, nicotine use, and relaxing. A timeline-based RS design method was proposed and browsing preference was added as an adjustment method.

## 3. Methodology

Following Adomavicius and Tuzhilin [21], a recommendation problem can be formulized as follows: Let $C$ denote the user set, $S$ be the recommendation items, and $u$ be a utility function that measures the usefulness of item $s$ to user $c$, i.e., $u: C \times S \to R$, where $R$ is a completely ordered set (e.g., nonnegative integers or real numbers within a certain range). Then, the recommendation problem can be modeled as a maximization formula:

$$\forall c \in C, s'_c = \arg\max_{s \in S} u(c,s). \quad (1)$$

For general recommendation such as products[24] or movies[25], the utility function always refers to how much a user likes a particular item. The common methods are CB, in which the user will be recommended items similar to those previously preferred by the user, CF, in which the user will be recommended items that similar users previously preferred, and the hybrid method, which combines CB, CF or others. These methods achieved great success in several recommendation scenarios and still saw wide use in health-aware domains [1,29,51,61].

However, for health-aware recommendation, these methods are insufficient, as more domain-related constraints should be considered. The usage of methods in HRS are shown in Table 2. In the next section, we discuss these methods.

Table2. The methods using frequency statistics in HRS in the last decade

| | 2010 | 2011 | 2012 | 2014 | 2015 | 2016 | 2017 | 2018&2019 |
|---|---|---|---|---|---|---|---|---|
| CF | 2 | 2 | 1 | 2 | 1 | 2 | 6 | 4 |
| CB | 2 | 2 | 0 | 0 | 1 | 1 | 3 | 0 |
| Hybrid | 1 | 0 | 0 | 0 | 1 | 1 | 0 | 0 |
| Demographic | 0 | 0 | 0 | 0 | 1 | 0 | 2 | 1 |
| KB | 3 | 0 | 3 | 0 | 1 | 1 | 5 | 4 |

The abscissa denotes the year, and the ordinate denotes how often the method is used in research articles.

### 3.1 Filtering methods

Filtering algorithms were broadly classified in to (a) content-based filtering; b) collaborative filtering; c) demographic filtering; and d) hybrid filtering [20]. We find that in this categorization method, the core ideology uses "similarity" measures to filter items for recommendation.

*3.1.1 Content-based filtering*

The content-based filtering method utilizes cumulative preference or rating data created by a user to predict the most likeable items. Details, and items are represented in several dimensions, e.g., the main features of items, such as ingredients [51], cuisine, preparation time [50] of a recipe, context features [47], nutrition [61,99] of a food, exercise type [29] of an exercise, or semantic features [71] of a personal health record (PHR). Then, a user profile can be modeled by the items he/she has preferred. A similarity measure such as cosine similarity [100] is always used to rate the relevance between an item and the user profile. Information retrieval technology has shown its abilities in the filtering process when the user profile has been built [101]. The content-based filtering method has its root in information retrieval (IR) [21,102,103]. Usually, a weighted scheme is used to differentiate the importance of item features. The most popular scheme is the term frequency/inverse document frequency (TF-IDF) [104,105]. In another direction, machine learning methods such as classification algorithms are also used to model the user profile [100]. They treat the problem of new item prediction as a type-tagging process such as like or dislike based on user tag data. For instance, nearest neighbor methods were used to cluster the dataset (here textual descriptions of implicitly or explicitly labeled items) into several types [104], and an item can be labeled from its nearest neighbors. Euclidean distance measure is a popular method to evaluate item similarity for clustering [50,63,104].

For a health recommendation, Achananuparp and Weber [47] used food diary context to model the content similarity for the recommendation. van Pinxteren et al. [50] applied Euclidean distance to compute the similarity of extracted feature representation of recipes. Ueda et al. [51] thought that the ingredients in recipes are the key features; the matching degree played its importance for recipe recommendation. The same work has Freyne and Berkovsky [1], who used ingredient ratings to evaluate the recipes. Kondylakis et al. [71] utilized the cosine measure to evaluate the correlation between user's query and document vectors. Hors-Fraile et al. [75] also used cosine to measure the similarity of the user interest vector and the item type vector. Bocanegra et al. [74] extracted the semantic features of SNOMED-CT terms to match the YouTube video title and Medline Plus links. Bianchini et al. [57] recommended food based on concept-based similarity such as the following:

$$Sim\left(\hat{r}_r, r_i\right) = \frac{2\sum_{c_r, c_i} \omega_r \cdot ConceptSim\left(c_r, c_i\right)}{\left|C_i\right|} \in [0,1], \qquad (2)$$

where $C_i$ denotes the concept feature sets of recipe $r_i$, $c_i$ ranges over set $C_i$, $c_r$ ranges over set $\hat{C}_r$, which denotes the concept feature sets of recipe $\hat{r}_r$, $\left|C_i\right|$ denotes the number of concepts in set $C_i$, and $\omega_r$ denotes the weight of concept $c_r \in \hat{C}_r$. However, the $ConceptSim(c_r, c_i)$ represents the concept similarity between $c_r$ and $c_i$ as follows:

$$ConceptSim(c_r, c_i) = \frac{2|c_r \cap c_i|}{|c_r| + |c_i|} \in [0,1], \quad (3)$$

where $|c_r \cap c_i|$ denotes the number of recipes that have both concepts, and $|c_r|$, $|c_i|$ denotes the number of recipes that have the corresponding concept.

*3.1.2 Collaborative filtering (CF)*

CF is the process of filtering or evaluating items through the opinions of other individuals [22,106]. CF relies on the individual ratings of all the users, and the preferences from similar users are recommended to each other. According to Breese et al. [23], algorithms for CF can be grouped into two classes: memory-based (or heuristic-based) and model-based. The memory-based algorithms are heuristics and can be formulated as follows [21]:

$$r_{c,s} = \underset{c' \in \hat{C}}{aggr}\, r_{c',s}. \quad (4)$$

Here, $r_{c,s}$ denotes the rating of current user $c$ of item $s$, and $\hat{C}$ denotes the set of $N$ users that are the most similar to $c$ and have rated item $s$. Examples of the aggregation function including the following:

$$\begin{aligned}
&(a)\ r_{c,s} = \frac{1}{N}\sum_{c' \in \hat{C}} r_{c',s}, \\
&(b)\ r_{c,s} = k\sum_{c' \in \hat{C}} sim(c,c') \times r_{c',s}, \quad (5) \\
&(c)\ r_{c,s} = \bar{r}_c + k\sum_{c' \in \hat{C}} sim(c,c') \times (r_{c',s} - \bar{r}_{c'}),
\end{aligned}$$

where k is usually selected as $k = 1 \big/ \sum_{c' \in \hat{C}} |sim(c,c')|$, and $\bar{r}_c$ is denoted as

$$\bar{r}_c = \left(1/|S_c|\right) \sum_{s \in S_c} r_{c,s}, \text{ where } S_c = \left\{ s \in S \,\middle|\, r_{c,s} \neq \varnothing \right\}. \quad (4)$$

Different approaches are used to measure $sim(c,c')$. The cosine similarity measure [22,107], and Pearson correlation [108,109] are common similarity measures where users are represented as vectors.

The model-based algorithms use the collection of user ratings to learn a model and then to make a prediction. The main difference between model-based and memory-based approaches is that the model-based techniques calculate utility (rating) predictions based not on ad hoc heuristic rules but on a model learned from the underlying data using statistical and machine learning techniques [21]. For a model-based system, linear regression, matrix factorization, Bayesian clustering, and associative classification can be used [46]. Here, we details the matrix factorization algorithms, which can be treated as a solver of data sparsity and cold-start problems (a problem of inability to address the system's new products and users[21,31,110]). In its basic form, matrix factorization characterizes both items and users by vectors of factors inferred from item rating patterns [110]. Matrix factorization models map both users and items to a joint latent factor space of dimensionality $f$ such that user-item interactions are modeled as inner products in that space. SVD is a famous matrix factorization technique that decomposes $m \times n$ matrix $A$ with rank $r$ into :

$$SVD(A) = U \times S \times V^T, \tag{5}$$

where $U, V$ are orthogonal matrices with dimensions $m \times m$ and $n \times n$. Using SVD in CF requires factoring the user-item rating matrix. This approach often suffers difficulties due to the sparsity of the user-item rating matrix. To learn the factor vectors $q_s \in R^f$ of item $s$ and $p_c \in R^f$ of user $c$, the system minimizes the regularized squared error on the set of known ratings as follows:

$$\min_{q^*,p^*} \sum_{(c,s)\in\kappa} (r_{cs} - q_s^T p_c)^2 + \lambda(\|q_s\|^2 + \|p_c\|^2). \tag{6}$$

Here, $\kappa$ is the set of $(c,s)$ pairs for which user-item rating $r_{c,s}$ is known, and $\lambda$ is the coefficient of regular terms. Additionally, the stochastic gradient descent (SGD) algorithm was adopted [111]. The algorithm loops through all the rating data in the training set for prediction $\hat{r}_{c,s} = q_s^T \cdot p_c$; then, current prediction error $e_{c,s} \overset{def}{=} r_{c,s} - \hat{r}_{c,s}$ is computed. The parameters are adapted by moving in the opposite direction of gradient, yielding the following:

$$q_s \leftarrow q_s + \gamma \cdot (e_{c,s} \cdot p_c - \lambda \cdot q_s), \tag{7}$$

$$p_c \leftarrow p_c + \gamma \cdot (e_{c,s} \cdot q_s - \lambda \cdot p_c), \tag{8}$$

where the $\gamma$ is a regulation proportion. If a bias of rating is considered $b_{c,s} = \mu + b_s + b_c$, the predicted item's rating is as follows:

$$\hat{r}_{c,s} = \mu + b_s + b_c + q_s^T \cdot p_c. \tag{9}$$

The minimization squared error function change to

$$\min_{b^*,q^*,p^*} \sum_{(c,s)\in\kappa} (r_{c,s} - \mu - b_s - b_c - q_s^T \cdot p_c)^2 + \lambda(\|p_c\|^2 + \|q_s\|^2 + b_c^2 + b_s^2). \tag{10}$$

Ornab et al. [46] used an item-item memory-based CF for which the similarity values between items are found and the most similar item is suggested to the user. The item vector was constructed by users' rating of it; if the number of users is "n", then the representation of an item will be an n-dimensional vector. Cosine similarity of two items was measured, and a similar meal was recommended. In addition, they performed the matrix factorization model, and item rating prediction was done via a dot product of user and meal vectors. Zhang et al. [68] used a user-doctor matrix to learn the latent factors of user and doctor. They considered the following rating offsets

$$RS_{ij} = \rho R_{ij} + (1-\rho) S_{ij}, \tag{11}$$

in which $RS_{ij}$ represents the revised rating on doctor $j$ by user $i$, $R_{ij}$ represent the original rating, $S_{ij}$ represents the emotional offset, and $\rho$ is used for adjusting the weight between original rating and emotional offset. Zhang et al. [72] proposed a "User-Item-Tag" three tuple considering the scoring of the user on drug and drug tag. Additionally, they used a new element that represents the predicted rating that the users give to the drugs according to certain tags and utilize the two dimensions of "User-Tag" in the three-order tensor to recommend the top-N drugs that have the highest scores.

Other model-based CF methods include SVM [6], which uses ingredient features to represent recipes, naïve Bayes [29], Bayesian network [67], nearest neighbor method [49], random forest [65], pairwise approaches [54-56,89,112], which used pairwise similarity to find recommendations.

For CB or CF filtering methods, similarity algorithm both acts as an important measure for new item prediction. van Pinxteren et al. [50] focused on the problem of alternative recipes. They extracted 55 features, including meal type (soup, salad, and hotchpotch) and cuisine (Dutch, and Italian, etc.) to represent recipes and measure recipe similarity by Euclidean distance. The recommendation was conducted based on the similarity measure and participants' past meal choices. Forbes and Zhu [49] applied a matrix factorization method to compute ingredient similarity based on user scores of recipes; the method shows advantages when a similarity measurement is compared to an appearing frequency method. Achananuparp and Weber [47] considered the food substitution problem in recipes. They used a food-context matrix to compute the cosine similarity between row vectors. Additional, for the data sparsity problem of the matrix, truncated singular value decomposition (SVD) was proposed for product dot produce similarity. Ornab et al [46] conducted a food recommender system on two CF strategies; the first is a memory-based vector similarity method that actually applies a user historical score from 1-5 on meal sets to represent items, and the cosine similarity measure was used to compute the most similar item. The model-based method was to factorize the user-item rating matrix into an individual user matrix and an item matrix by alternating least square algorithm [113], and prediction was conducted via the two matrixes.

*3.1.3 Demographic filtering*

The demographic filtering method is justified on the principle that individuals with certain common personal attributes (gender, age, and country, etc.) should have common preferences [20]. That is, two users could be considered similar not only if they rated the same items but also the demographic segment should function. Demographic filtering was considered an extension of traditional collaborative filtering [21,114]. Demographic filtering was widely used in health recommendations, and the technique was considered a supplement for the problem of user rating sparsity, which usually occurs in health-related RSs. Age and gender as basic attributes are widely used in demographic filtering-based recommendations. Jabeen et al. [65] indicated that the standard collaborative filtering algorithm is not suitable for medically related problems, especially CVD. They used a community-based recommender system with three steps — gender, age, and patient similarity computed by remaining features—for CVD advice recommendations. Rehman et al. [58] attempted to utilize user gender, age and the value of pathological test reposts to compare with the normal ranges. This last value was used to produce nutritional restrictions for recommending foods for the user. In Jabeen et al. [65], physical and diet plan recommendations to a CVD patient were produced after classifying the user into age and gender groups. Agu and Claypool [29] integrated player age and gender into exercise game recommendations. Hors-Fraile et al. [75] used attributes such as gender, employment status, age, and quitting date to calculate user similarity. Elsweiler and Harvey [3,61] incorporate the user's personal personas such as age, gender, height, etc., into a Harris-Benedict equation [115] to calculate the individual's BMR and daily calorie requirements, and then calculate a meal recommendation.

### 3.1.4 Hybrid

The hybrid method is a combination of different filtering approaches; it was created to solve the limitations of a single method [116]. Due to the complexity of health domains, that recommendation should not always consider user preference, but healthiness plays a more important role; a hybrid recommendation is widely used in health-related recommender systems for this purpose. There were many hybrid methods, such as content and collaborative [1,48,75], content-demographic [58], collaborative-demographic [61,65,69], knowledge-content [57,63], and knowledge-collaborative [54,60,97]. The hybrid method is increasingly attracting attention, as it functions well in health domains.

## *3.2 Knowledge-based recommendation*

Recommender systems that use additional knowledge sources are defined as KB because they relied more heavily on knowledge sources than on techniques [77]. A knowledge-based recommendation attempts to suggest objects based on inferences about a user's needs and preferences [117]. It always considers the "need" problem for a user, whereas to some extent, traditional recommendation approaches are more used to concern the "like" problem. Compared to content-based and collaborative filtering approaches, the knowledge-based approach has no cold-start problem [118]. However, it also suffers from the so-called knowledge acquisition bottleneck in which knowledge engineers must work hard to convert the expert domain knowledge into formal, executable representations [118].

There are two well-known knowledge-based recommenders. One is case-based reasoning in which items are identified as cases and the recommendation is generated by retrieving the most similar case to the user's query or profile [119,120]. Another is a constraint-based recommender, which takes into account explicitly defined constraints (e.g., filtering constraints or incompatibility constraints) [77].

### *3.2.1 Case-based reasoning*

For case-based reasoning, Torrent-Fontbona and López [17] used case-based reasoning to adapt a past bolus insulin recommendation to the current state of the user, retaining updated historical patient information to deal with slow and gradual changes in the patient over time. Given a problem or query Q, case-based reasoning obtains the result in four steps: 1) retrieve past experiences similar to Q, 2) reuse of retrieved past solutions to work out a new solution for the new problem, 3) revision of the proposed solution according to the outcome, and 4) retention of the new case according to a strategy that considers its relevance measured according to certain metrics. A case was represented by the input of the time of day, the estimated carbohydrate intake, the current blood glucose level, and past and future physical activity, and outputs of insulin to carbohydrate ratio (ICR) and the insulin sensitivity factor (ISF), which were key values for insulin bolus calculation. A retrieval step was conducted to select similar cases to the query by using a similarity measure inspired by Tversky index [121] and combining an average Euclidean distance between all the features. In the reuse step, ICR and ISF were separately computed by using the case similarity, and then the bolus recommendation was calculated.

### *3.2.2 Constraint-based recommendation*

For a constraint-based recommendation such as a food recommendation, nutritional restrictions for a normal individual [54-57,61,122], and diet restrictions for disease [34,53,59,62,63] are two basic constraint types. The sources from which constraints collected are mainly existing guidelines from authoritative organizations such as the WHO or government departments [54,56], domain expert knowledge [14,16,57,59,62,63,76], classical computational formulas that are usually applied with demographic information to produce individual constraints [34,53,61,62], or even statistical laws from large users' data [60,97]. Ontology, as a formal knowledge representation method functioning in constraint-based recommendations, represents the domain concepts and relationships between concepts [123]. The concepts cover the content of personal profile, nutrition, food, disease, health constraints, drug, and testing, etc., [16,34,57], and relationships have "is-part-of", "association" [62], "subclass-of", "composed-of", "constrained-by" [57], and "has"[59]. In terms of an ontology development tool, Protégé[1] is a prevalent platform created by the Stanford Center for Biomedical Informatics Research for the web ontology language (OWL)-based ontology development [34,124]. Protégé has been widely used in developing medically related ontology studies [16,125]. The individual constraints and recommendation are generated using inference techniques such as rule-based [12,16] and conditional probability distribution [97] and tools such as JENA [57,59], which provides an API (application program interface) to extract data from and write to resource description framework (RDF) graphs, and JESS [16], which is a rule engine and scripting environment written entirely in java, Pellet and FACT reasoners [126].

Occasionally, the classical logic of ontology such as true or false cannot describe real world concepts such as imprecise and vague information [62,127]. For example, in classical logic, if the weight of a user changes 1 kg, the user may possibly change from normal to overweight; then, the recommendation would be completely different. Fuzzy logic is an available method to solve this problem and has been applied to analyze the daily requirements for calories, fat, protein, and carbohydrates for a patient with chronic diseases. Lee et al. [62] used Type-2 fuzzy logic to generate personalized diabetes diet recommended applications. They used a Type-2 fuzzy set to represent concepts in food and personal profiles. A type-2 fuzzy food ontology containing four concepts—"servings of six food groups each portion", "nutrition facts each portion", "contained calories of each serving of six food groups", and "contained calories of nutrition each gram"—and a fuzzy personal food ontology were built. Then, a fuzzy-based inference mechanism was implemented to recommend a dinner allowance based on obtained calories at breakfast and lunch.

### *3.3 Group recommendation*

Health recommendations also show a characteristic of being "group-based". There are many scenarios in which health recommendations are better for recommending to a group rather than to individuals [128]. For example, for food recommendations, a family-based diet consumption model is a common scenario around the world. No matter whether a food plan is recommended to a normal family or to a "disease" family (which refers to a family that contains patient(s) such as diabetics), considering the family as a whole will achieve a more acceptable recommendation and facilitate the dinner maker as far as possible. Group-based recommendations can be categorized according to two standards: how individual preference is obtained, and whether recommendations or profiles are aggregated [128]. The first standard differentiates the content-based and collaborative filtering recommendations. The second standard considers two cases: recommendations for individuals and aggregated for a group, and individual preferences that are aggregated into a group model and generate group-based recommendations. Aggregation strategies have plurality voting which uses "first past the post", average for averages individual ratings, and multiplicative for multiple individual ratings, etc. [128,129].

[1] Available: https://protege.stanford.edu/

Berkovsky and Freyne [52] compared the two aggregation strategies of aggregated models and aggregated predictions. The main difference between the two strategies is whether user models are aggregated into a group for recipe recommendations or individual predictions are aggregated into a group recipe recommendation. They used a weighted model to simulate the individual impacts in a family such as the "most respected person" aggregation strategy. Additionally, a hybrid model was proposed to show the possibility of merging individual recommendations with aggregated ones. The evaluation shows that the personalized strategy for individual recommendations (recommend recipes for individuals by CF) achieved the highest accuracy but lowest coverage, while the group-based strategy was moderate, and a general strategy (recommend the most popular items) achieved the lowest accuracy but highest coverage, which activated the authors to hybridize these strategies according to the density of the user data.

## 4. Evaluation

Evaluation of predictions and recommendations has shown their importance since the RS research was created [44]. There are three types of evaluation experiments: 1) offline, which uses existing data sets and a protocol that models user behavior to estimate recommender performance with measures such as accuracy; 2) user study, which employs with real user interactions with the RS; and 3) online, which measures the change in user behavior when interacting with different recommendation systems [130]. For a traditional RS, the purpose of a recommendation is to recommend items that the user would like. For this result, benchmarks such as the open source toolkit RiVal[2] [131] for RS evaluation were made, and open datasets Netflix [28], Amazon music[3], Movielens[4], Jester[5], LibRec[6], Dook-Crossing[7], etc. and algorithms were conducted and compared to improve recommendation quality and effectiveness [22,91,106,132,133]. The commonly used evaluation scenarios are the following: (1) prediction evaluations, (2) evaluations for recommendation as sets, and (3) evaluations for recommendations as ranked lists [20,44,134]. Additionally, according to the different measures, corresponding evaluation metrics are made. For measuring prediction, accuracy and coverage are common metrics. Accuracy metrics include mean absolute error (MAE) [133], root of mean square error (RMSE) [55], normalized MAE, normalized RMSE, average MAE, and average RMSE[130]. The MAE and RMAE formulas are as follows:

$$MAE = \sqrt{\frac{1}{|\mathrm{T}|}\sum_{(u,i)\in \mathrm{T}} |\hat{r}_{ui} - r_{ui}|},$$

$$RMSE = \sqrt{\frac{1}{|\mathrm{T}|}\sum_{(u,i)\in \mathrm{T}} (\hat{r}_{ui} - r_{ui})^2},$$

[2] https://github.com/recommenders/rival
[3] http://jmcauley.ucsd.edu/data/amazon/
[4] https://movielens.org/
[5] https://goldberg.berkeley.edu/jester-data/
[6] https://www.librec.net/datasets.html
[7] http://www2.informatik.uni-freiburg.de/~cziegler/BX/

where T is the test set of user-item pairs $(u,i)$, $\hat{r}_{ui}$ is the predicted rating, and $r_{ui}$ is the true rating. Normalized and average RMSE are the extended versions. Coverage often refers to the proportion of items that the RS can recommend.

While a recommendation might be an item set such as the scenario of recommending movies of interest to a user which he will add to his "to watch" queue, the recommender was not evaluated on how properly predictions are of the ratings of items but rather on how many movies in the recommendation set will be added to the queue by the user. The most widely used recommendation quality measures are precision, recall, and F1:

$$Precision = \frac{tp}{tp + fp},$$

$$\mathrm{Re}call = \frac{tp}{tp + fn},$$

$$F_1 = \frac{2 \times Precision \times \mathrm{Re}call}{Precision + \mathrm{Re}call},$$

where $tp$, $fp$, $fn$ refer to the recommendations which the user accepted, or did not accept, and the used items which the RS did not recommend. Additionally, receiver operating characteristic (ROC) curves with true and false positives were also applied for measuring the quality of recommendation acceptance.

When recommending a list of items with order, their rank shows importance. For example, for restaurant recommendations, those higher in the order should be more relevant to the user's taste. There is no need to rate the restaurants; however, this approach makes no sense with a recommendation of a restaurant set. The common ranking measures are the following: R-score metric, and normalized cumulative discounted gain (NDCG) [135]. The R-score formula is as follows:

$$R_u = \sum_u \sum_j \frac{\max(r_{ui_j} - d, 0)}{2^{(j-1)/(\alpha-1)}}$$

where $i_j$ refers to the item ranked in the jth position, $r_{ui}$ is user $u$'s rating of item $i$, $d$ is a task-dependent neutral rating, and $\alpha$ is a half-life parameter. NDCG is a measure of information retrieval, which is based on cumulated gain $g_{ui}$ for user $u$ being recommended an item $i$. The average (DCG) for a list of $J$ items is defined as follows:

$$DCG = \frac{1}{N} \sum_{u=1}^{N} \sum_{j=1}^{J} \frac{g_{ui_j}}{\max(1, \log_b j)}.$$

Then, NDCG is defined as follows:

$$NDCG = \frac{DCG}{DCG^*}$$

where $DCG^*$ refers to DCG calculated on the ideal item ranking.

Other recommendation evaluation measures have confidence, which refers to the system's trust in its recommendation or prediction [136]; novelty, which represents whether the RS recommendation is a novel item for users; diversity, which refers to how diverse is the recommendation set; utility, which refers to the improvement caused by the RS; risk, which refers to the risk the RS may bring; privacy, which refers to how the RS maintains a user's privacy preferences and does not do other things with these data; and stability, which refers to how the RS does not change significantly in a short time, etc.

For the evaluation of an HRS, the accuracy evaluation metrics of a traditional RS are necessarily considered. The three evaluation scenarios of prediction [1,48,49,52,60], recommend a set [16,60,72], and recommend a rank list [47,58,74] all exist in an HRS. Accuracy measurement metrics such as MAE [55,65], RMSE [48,49], precision, recall, and F1 [72] were used. Additionally, there are famous domain datasets such as Allrecipes.com[8], which is one of the most popular recipe-sharing website for recipe recommendation [6,50]; Yelp[9], which has user's reviews and ratings for restaurants, dentists and doctors [68]; ICD9-CM[10], which is for diseases [53]; and SNOMED-CT[11], which is for clinical concepts and relationships [74]. Occasionally, the prediction, set recommendation and ranking list scenarios are combined, and related metrics are collaborative to evaluate the entire RS. Berkovsky and Freyne [52] gathered a dataset of explicit symbolic ratings on a 5-point Likert scale range from hate to love from participants for a recipe corpus. MAE was used to measure the rating prediction strategies, and the F1 metric was used to measure the recipe recommendation set offline. Jabeen et al [65] collected 100 cardiac patients who visit the hospital for their routine checkup at a coronary artery unit. The dataset was labeled by an expert cardiologist, and the cardiovascular disease advice recommendation was evaluated by MAE to identify the variation with actual recommendations by a cardiologist. Gräßer et al [69] combined a therapy response prediction with the reliable ranked therapies, so they used RMSE to quantify the difference between estimated response and real response and Precision@N for ranked list evaluation.

In addition to accuracy, other evaluation measures were conducted. Trattner and Elsweiler [54] applied WHO and United Kingdom FSA health criteria to score the recipes and then evaluate the *healthiness* of recipe recommendations. In [60], *satisfaction* was evaluated by a user study; 100 obese youth were gathered and experienced the nutrition RS for 90 days. Then, the subjects were asked to score the service satisfaction level on a 5-point scale. Ge et al. [48] applied the system usability scale, which is a widely accepted measurement for evaluating system *usability,* with 20 users and then evaluated the quality of their food RS. The interaction style with users was using a questionnaire with 10 standard questions, in which a Likert scale ranging from 1-5 (1 for disagree, 5 for agree) was set. van Pinxteren et al. [50] invited a small set of participants to actually cook recommended recipes to evaluate the *effectiveness* of their tailored recipe recommendation. Then, a satisfaction metric was conducted via a 7-point Likert scale. Torrent-Fontbona and López [17] evaluated the effectiveness on 33 virtual subjects for a 90-day simulation with their bolus recommendation for type 1 diabetes. Times in, below and above the glycemic target were compared with comparable systems. Additionally, *robustness* was evaluated when the system was missing 10% of input values such as physical activity. *Coverage* was considered in [1,52]. Yang et al. [55] evaluated the *acceptance* of their meal RS with a user study on 60 participants; similarly, Phanich et al. [63] invited nutritionists to design a questionnaire for food RS acceptance evaluation. System *response time* to generate menus was evaluated in [57]. Bielik et al. [14] tested the *understanding* of their physical active recommender system with a group of 12 children. Bauer et al. [76] used a 4-week user study to explore whether their HRS could promote *awareness* about sleep recommendations.

[8] https://www.allrecipes.com/
[9] https://www.yelp.com/dataset
[10] https://www.cdc.gov/nchs/icd/icd9.htm
[11] https://www.snomed.org/

## 5. Challenge and trends

HRS has had a short-term development; when searching with "health recommender system" in web of science, the oldest article is from 2007 [31]. There are many constraints to challenge researchers developing a health-related RS. As a special RS, an HRS suffers the challenges of RSs. There is evidence showing that the challenges for an RS include aspects such as explanation, sparsity, available item limitations, new users, and cold-start [69,72,137]. More challenges for an RS can be seen in [20,79]. Here, we summarized some exclusive challenges for HRS that should be considered for future studies.

The first key challenge is how to keep the recommendation "healthy", which should be regarded as the vital factor of an HRS. Here, healthy indicates that the system should not only take preference into account but also give more attention to user profiles such as diseases, symptoms, constraints, etc., and then make tailored recommendations. Health considerations affect the users, especially patients' acceptance of the recommendation results, degree of expert approval, recommendation effectiveness, and satisfaction, etc. Almost all the evaluation measures of an HRS should be built under the health dimension. However, how can we make an HRS "Healthy"? Current strategies have 1) directly involved experts such as doctors [71] and nutritionists [63] in the recommending process, 2) used domain data such as PHR [67] and prescriptions [57], which are based on expert knowledge, and 3) take advantage of domain guidelines and knowledge for user profiling [53] and health inferences [57]. Directly engaging experts would be easy to conduct but problems exist. A single expert would not be enough for the whole RS designation. Then, developers would produce a biased RS under a single's expert knowledge. More than one expert would save this awkward situation with respect to discussions, but occasionally conflicts would arise between them in which it would not be easy to obtain a final agreement. Additional, using personalized data such as PHR would suffer privacy and accessibility problems. When integrating domain guidelines and knowledge into an RS, one should first consider how to represent this knowledge for computation—for example, for similarity measuring or logic inferring, etc.

Another key challenge is how to evaluate the "real effects" of the recommendation. For an RS-like product recommender, the effect would be simply evaluated by whether the user buys. However, the health recommendation might lead to long-term effectiveness. For example, a food recommendation for diabetes may not cause high blood glucose in the glucose measure dimension, but damage might possibly accumulate and cause implicit impacts in some other dimension. Offline evaluation is widely used for system accuracy measuring, which was considered a measure of the algorithm on whether it can predict or rank standard recommendations. It is a measure of the algorithm of ability, not an "effect" measurement of the recommendation. For this purpose, user studies are conducted; real users are collected for usefulness testing. However, a small-scale number of participants in a short-term experiment is not enough for the "real effect" evaluation. Many authors mentioned that further experiments should be carried out [14,70,76,138].

Other challenges exist, such as a lack of education causing patients to reject using an HRS [45], ethical concerns [99], culture differences [139], financial problems [31].

There is has no doubt that RSs should play their role in the health domain. The development of HRSs shows its power in solving disease management and lifestyle habit improvement issues. The trend of HRS studies is to adapt this technique to solve more scenarios and make efforts to find solutions for these challenges. As further studies continue, more problems will be discovered and solved to build adaptive systems.

## 6. Conclusion

In this survey, we summarized the highly influential works in health recommendation systems from the past ten years. We reviewed the main aspects of interests, methodology, evaluations, challenges and trends. We found that health recommendation research interests mainly concern diet, physical exercise, medically related, and other obscure issues. Diet recommendations attract the most popular attention due to their involving more-accessible data, guidelines, formulas, and less risk. A diet recommendation should be a simple operable work in the HRS research. For a recommendation method, a traditional filtering method such as content-based and collaborative filtering shows its ability in HRS but not enough to address health constraints. The knowledge-based recommendation method functions well, as a constraint-based knowledge recommendation can solve the health constraint involvement problem and cold-start, and new user, etc., problems. For the evaluation of an HRS, the most important factor is safety for users. To ensure safety, various evaluation measures in addition to accuracy were proposed such as healthiness, usability, and robustness. However, evaluating an HRS is a multi-dimension work and more complicated than evaluating an RS. Future work should be focused on the specific challenges such as how to keep a recommendation healthy, and how to evaluate the real effect of an HRS. This investigation is a summary of recent HRS studies; we want to provide domain researchers some enlightenment and cautions for future works to implement HRSs in real life.

**Acknowledgements**

We would like to thank the nutritional department of the 1st Affiliated Hospital of Harbin Medical University for proving consultation. We also want to thank the anonymous reviewers for their comments, which provided us with significant guidance.

**Funding**

This research was financially supported by the Open Research Fund from Shenzhen Research Institute of Big Data, under Grant No.2019ORF01011

**Role of the funding source**

The funding source has played a role in this study design; data collection, analysis; paper publishing.

**Declarations of interest**

none

**Data availability**

none